\documentstyle[multicol,aps,prb,epsf]{revtex}

\begin{document}

\draft

\title{Superconductivity induced by inter-band nesting \\
in the three-dimensional honeycomb lattice}

\author{
Seiichiro Onari, Kazuhiko Kuroki$^1$,Ryotaro Arita, and Hideo Aoki
}

\address{Department of Physics, University of Tokyo, Hongo,
Tokyo 113-0033, Japan}
\address{$^1$Department of Applied Physics and Chemistry,
University of Electro-Communications, Chofu, Tokyo 182-8585, Japan}

\date{\today}

\maketitle

\begin{abstract}
In order to study whether the inter-band nesting can favor superconductivity 
arising from electron-electron repulsion in a three-dimensional system, 
we have looked at the repulsive Hubbard model on 
a stack of honeycomb (i.e., non-Bravais) 
lattices with the FLEX method, partly 
motivated by the superconductivity observed in MgB$_2$.  
By systematically 
changing the shape of Fermi surface with varied band filling $n$ 
and the third-direction hopping, we have found that 
the pair scattering across the two-bands is indeed found to 
give rise to gap functions that change sign across the bands 
and behave as an s- or d-wave within each band.  
This implies (a) the electron repulsion can 
assist {\it gapful} pairing when a phonon-mechanism 
pairing exists, and 
(b) the electron repulsion alone, when strong enough, can give rise to 
a d-wave-like pairing, which should be, for a group-theoretic 
reason, a time-reversal broken d$+$id with {\it point nodes} in the gap.
\end{abstract}

\medskip


\begin{multicols}{2}
\narrowtext

\section{introduction}
Recent discovery of the superconductivity in MgB$_2$\cite{akimitsu} 
with relatively high transition temperature ($T_c \sim 39$K) has 
invoked renewed interests in $sp$-bonded materials.
Electronically, the system is a $\pi$ electron system on layered 
honeycomb lattice, which immediately reminds us of  
graphite intercalation compounds (GIC)
such as LiC$_6$\cite{holzwarth} or KC$_8$\cite{inoshita,divincenzo}.

While the GIC is considered to be a conventional superconductor 
with $T_c <5$K, MgB$_2$ has an unusually high $T_c$ for 
$sp$-bonded materials (with a recent exception of 
C$_{60}$-FET structure\cite{batlogg}).  
Recently, Choi {\it et al.}\cite{Louie} have 
used an ab-initio pseudopotential density 
functional theory to solve Eliashberg's equation numerically, 
and have reproduced $T_c\sim 39$K, isotope-effect 
exponent $\alpha_B\sim 0.3$\cite{hinds,bud'ko}, and have
obtained a gapful BCS pairing, which is consistent with experimental 
results such as specific heat\cite{wang,bouquet}, 
tunneling and photoemission 
spectra\cite{takahashi,tsuda,karapetrov,szabo}, 
penetration depth\cite{manzano}, and the Raman spectra\cite{chen}. 
Thus GIC and MgB$_2$ both seem to be mainly phonon-mediated 
superconductors.  
However, to realize a high $T_C$, electron 
repulsion should not stand in the way of the phonon mechanism, 
so the question may be paraphrased: can the electron repulsion 
stand away from or possibly even assist the phonon-mediated pairing.  

On a more positive side, 
superconductivity from electron-electron repulsion itself is
fascinating in many ways, but there are many open questions.  
While there is a growing consensus that high-$T_C$ cuprates 
may be related to the electron correlation, we are only 
beginning to understand the link between the underlying 
band structure and the way in which the electron-mechanism 
superconductivity appears\cite{arita}.  Indeed, 
the way in which the superconductivity occurs is sensitively 
affected by the shape of the Fermi surface.  
Recently, two of the present authors proposed\cite{kuroki} that 
multi-band systems should open a new possibility of 
much higher $T_c$, where 
a fully gapped BCS gap function can appear when the 
Fermi surface consists of disconnected pieces, 
while the usual wisdom dictates that the repulsion-originated 
superconductivity should have, as in the cuprates, a strongly 
anisotropic gap function with nodes.  
They have conceived and demonstrated that, for 
some two-dimensional lattice models, the pair scattering 
(the Coulombic matrix elements that scatter pairs of 
electrons across the Fermi surface) can occur across the 
pockets, which gives rise to a gap function that changes sign across
the pockets with the same sign within each pocket.

So the purpose of the present paper is a 
combination of the above two motivations.  Namely, 
we study whether the {\it inter-band nesting} can favor superconductivity 
arising from electron repulsion in a three-dimensional system, 
by taking the repulsive Hubbard model on 
a stack of honeycomb lattices as a prototype.  
There, the honeycomb, a typical non-Bravais lattice, 
provides two pieces of the Fermi surface 
arising from the two bands, while the stacking can provide a 
natural nesting along that direction.  
So, if the pair scattering across the two bands 
along the nesting vector works favorably, 
we can expect a pairing from the interband nesting 
with the BCS gap with opposite signs across the two bands.

So we have studied the systematic dependence of the 
solution of Eliashberg's equation with 
the multi-band fluctuation exchange approximation on 
the shape of Fermi surface by changing the band filling 
and the third-direction hopping.  
The presence of a strong inter-band nesting is found to 
indeed give rise to gap functions that change sign 
across the two bands, but, if we turn to the symmetry 
within each band, there exist two, 
nearly degenerate modes that behave respectively 
like s- and d-waves.  
This implies (a) the electron repulsion can assist an s-wave like
 pairing when the phonon-mechanism pairing exists as a dominant
 mechanism.  We have further found 
that (b) when the electron repulsion is strong enough, 
the d-wave is realized, which should be, for a group-theoretic 
reason, a time-reversal broken linear combination of 
two symmetries with {\it point} nodes in the gap.

According to the band calculation\cite{kortus},
the Fermi surface of MgB$_2$ consists 
of two tubular networks having a boron-$2p\pi$ character along with
two cylinders of the boron-$2p\sigma$ characters.
As has been stressed by several authors\cite{yamaji,furukawa},
the nesting between these two $\pi$ bands are quite good.  
Although $p\sigma$ bands are considered to be important 
in that the electron-phonon interaction is much stronger in this 
band, here we concentrate on the $p\pi$ bands in order to 
focus on the effects of the Fermi surface nesting.
The GIC, on the other hand, has a much smaller 
inter-layer transfer energy ($t_z$), so that 
the Fermi surface is a cylinder of 
carbon $\pi$ character with no dominant nesting vectors. So in 
the latter part of the paper we shall cover both MgB$_2$- and 
GIC-situations in a systematic variation of $t_z$ 
and band filling. 
The systematic study also serves to explore how 
the inter- and intra-band 
nesting compete in realizing superconductivity.

\section{formulation}
\begin{figure}
\begin{center}
\leavevmode\epsfysize=25mm 
\epsfbox{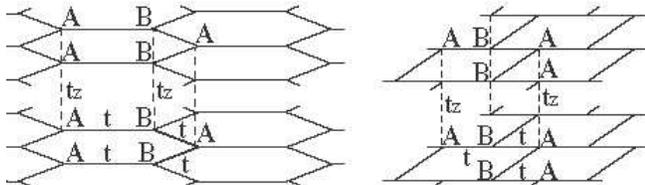}
\caption{3D layered honeycomb lattice (the left panel), which is 
topologically equivalent to the lattice in the right panel.  
A and B indicate sublattices.}
\label{lattice}
\end{center}
\end{figure}

Now let us start with the case corresponding to MgB$_2$.  
Spin-fluctuation-mediated superconductivity is here studied with 
the fluctuation exchange approximation (FLEX) developed by 
Bickers {\it et al.}\cite{bickers1,bickers2,dahm,bennemann}.  
The model is a 3D two-band Hubbard model with the repulsion $U$ 
on layered honeycomb lattice,
\begin{eqnarray}
{\cal H}&=&\sum_{\langle i,j\rangle,\sigma}
\sum_{\alpha,\beta}^{\rm A,B}t_{ij}\left(c_{i\sigma}^{\alpha\dagger}c_{j\sigma}^{\beta}+{\rm H.c.}\right)\nonumber\\
& &+ U\sum_{i}\sum_{\alpha}n_{i\uparrow}^{\alpha}n_{i\downarrow}^{\alpha},
\end{eqnarray}
where $U$ is the Hubbard repulsion.  The essential ingredient here is the 
non-Bravais lattice having A and B sublattices, 
which has two bands within a layer.  
Here we assume a non-staggered layer 
stacking as realized in MgB$_2$ and GIC 
(Fig.\ref{lattice}), which also depicts the
 inter-layer hopping $t_z$. Hereafter we take 
the intra-layer hopping $t=-1$.

The noninteracting band dispersion for 3D honeycomb is
\begin{equation}
\epsilon({\bf k})=2t_z\cos k_z\pm t\sqrt{3+2\left[\cos k_x+
\cos k_y+\cos(k_x+k_y)\right]}.
\end{equation}

In the two-band FLEX\cite{koikegami,kontani}, 
Green's function $G$, self-energy $\Sigma$, 
spin susceptibility $\chi$, and the gap function
$\phi$ all become $2 \times 2$ matrices, such as $G_{\alpha\beta}(k)$,
where $k\equiv({\bf k},i\omega_n)$ with $\omega_n=(2n-1)\pi T$ being the
Matsubara frequency for fermions.

The self-energy is given by
\begin{equation}
\Sigma^{(1)}_{\alpha\beta}(k)=\frac{T}{N}\sum_qG_{\alpha\beta}(k-q)V_{\alpha\beta}^{(1)}(q),\label{equbegin}
\end{equation}
where the fluctuation-exchange interaction $V^{(1)}(q)$ is
\begin{eqnarray}
V_{\alpha\beta}^{(1)}(q)&=&\frac{3}{2}U^2\!\!\left[\frac{\chi^{{\rm irr}}(q)}{1-U\chi^{{\rm irr}}(q)}\right]_{\alpha\beta}\!\!\!+\frac{1}{2}U^2\!\!\left[\frac{\chi^{{\rm irr}}(q)}{1+U\chi^{{\rm irr}}(q)}\right]_{\alpha\beta}\nonumber\\
&&-U^2\chi^{\rm irr}_{\alpha\beta}(q)
\end{eqnarray}
with
\begin{equation}
\chi_{\alpha\beta}^{{\rm irr}}(q)=-\frac{T}{N}\sum_{k}G_{\alpha\beta}(k+q)G_{\beta\alpha}(k).
\end{equation}
Here we denote $q\equiv({\bf q},i\epsilon_l)$  with $\epsilon_l=2\pi lT$ being 
the Matsubara frequency for bosons, and $N$ the number of ${\bf k}$-points 
on a mesh.

With Dyson's equation,
\begin{equation}
\left[G(k)^{-1}\right]_{\alpha\beta}=\left[G^{0}(k)^{-1}\right]_{\alpha\beta}+\Sigma_{\alpha\beta}(k),\label{equend}
\end{equation}
where $G^{0}$ is the bare Green's function 
$G_{\alpha\beta}^{0}(k) = 
[(i\omega_n+\mu-\epsilon_{\bf k}^{0})^{-1}]_{\alpha\beta}$
with $\epsilon_{\bf k}^{0}$ the bare energy, 
we have solved eqns.(\ref{equbegin})-(\ref{equend}) self-consistently.

$T_c$ may be obtained from Eliashberg's equation (for the 
spin-singlet pairing),
\begin{eqnarray}
\lambda\phi_{\alpha\beta}(k)&=&-\frac{T}{N}\sum_{k'}\sum_{\alpha',\beta'}\nonumber\\
& &V_{\alpha\beta}^{(2)}(k-k')G_{\alpha\alpha'}(k')G_{\beta\beta'}(-k')\phi_{\alpha'\beta'}(k')
\label{eliashberg},
\end{eqnarray}
where $\phi$ is the gap function, 
and the pairing interaction $V^{(2)}(k)$ is given as
\begin{eqnarray}
V_{\alpha\beta}^{(2)}(q)&=&\frac{3}{2}U^2\!\!\left[\frac{\chi^{{\rm irr}}(q)}{1-U\chi^{{\rm irr}}(q)}\right]_{\alpha\beta}\!\!\!-\frac{1}{2}U^2\!\!\left[\frac{\chi^{{\rm irr}}(q)}{1+U\chi^{{\rm irr}}(q)}\right]_{\alpha\beta}\nonumber\\
&&+U\delta_{\alpha\beta}.
\end{eqnarray}
$T_c$ is determined as the temperature at which 
maximum eigenvalue $\lambda$ becomes unity.

The susceptibility,
\begin{equation}
\chi_{\alpha\beta}({\bf k},0)=\left[\frac{\chi^{{\rm irr}}({\bf k},0)}{1-U\chi^{{\rm irr}}({\bf k},0)}\right]_{\alpha\beta},
\end{equation}
may be expressed as diagonalized components,
\begin{equation}
\chi_{\pm}=\frac{\chi_{{\rm AA}}+\chi_{{\rm BB}}}{2}\pm\sqrt{\left[\frac{\chi_{{\rm AA}}-\chi_{{\rm BB}}}{2}\right]^{2}+\left|\chi_{{\rm AB}}\right|^{2}} .
\end{equation}
Throughout this study, we take $N=32^3$ ${\bf k}$-point meshes, and the
Matsubara frequencies $\omega_n$ from $-(2N_c-1)\pi T$ to
$(2N_c-1)\pi T$ with $N_c=4096$, which gave converged results.

\section{result}
\subsection{MgB$_2$}
\begin{figure}
\begin{center}
\leavevmode\epsfysize=40mm 
\epsfbox{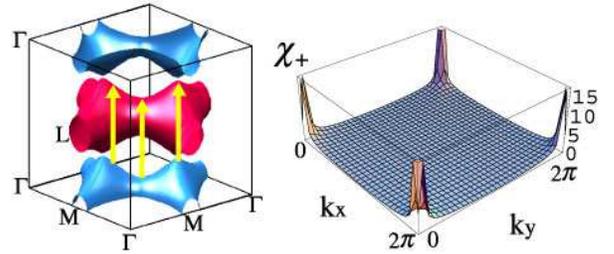}
\caption{Fermi surface (left panel, blue: bonding band, 
red: anti-bonding band) and $\chi_+$ (right panel) 
with $k_z=\pi$ for $U=1.5,n=1.03,t_z=0.65,T=0.01$. Yellow arrows
 indicate the nesting vector. } 
\label{MgB2-fermi-chi}
\end{center}
\end{figure}

Let us first discuss the spin structure.
We have fitted the shape of the Fermi surface to that 
obtained by the band calculation\cite{kortus} 
to have $t_z=0.65, n=1.03, U=1.5$ at $T=0.01$, where $n$ is band filling
($n=1$ for half filling)\cite{U=1.5}. 
The Fermi surface, Fig.\ref{MgB2-fermi-chi}, consists of two sets of tubular 
networks corresponding to the bonding and anti-bonding $\pi$ bands. 
The spin-susceptibility 
$\chi_+({\bf k},0)$ displayed in the same figure shows 
a sharp peak around $(0,0,\pi)$ reflecting 
a good nesting along the $z$ direction (arrows in Fig.\ref{MgB2-fermi-chi}).

\begin{figure}
\begin{center}
\leavevmode\epsfysize=120mm 
\epsfbox{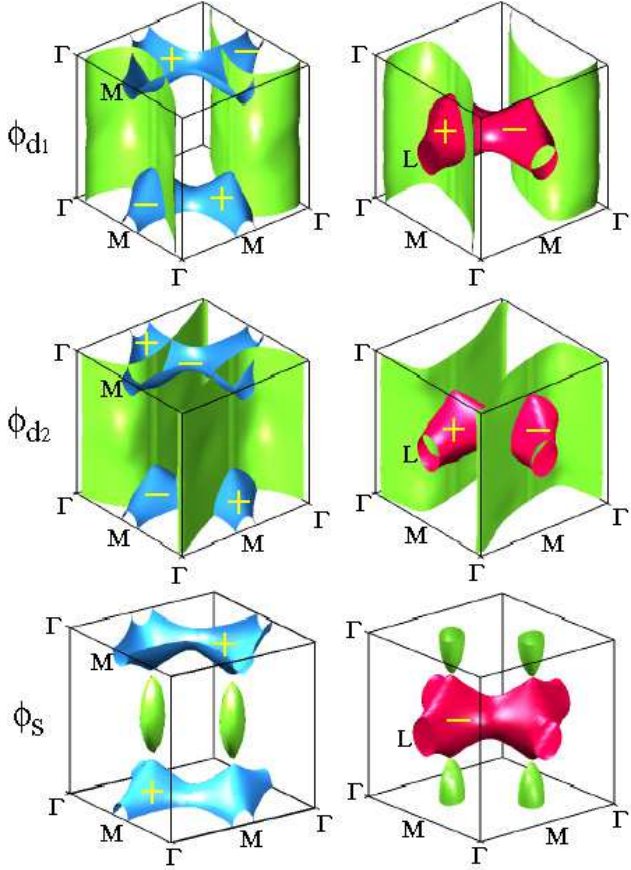}
\caption{The sign of the gap functions $\phi_{d_1}$(top), $\phi_{d_2}$(middle), 
and $\phi_s$(bottom) for  $U=1.5,n=1.03,t_z=0.65,T=0.01$.  
We have displayed the sign of the gap on the Fermi surface 
(left panels: bonding band, right panels: anti-bonding band) 
along with the nodal planes displayed in green.}
\label{MgB2-phi}
\end{center}
\end{figure}

Figure \ref{MgB2-phi} shows the gap function obtained from 
Eliashberg's
equation. The solutions having the largest $\lambda$ 
are gapless $\phi_{d_1}$ and $\phi_{d_2}$, 
which are degenerate.  
As expected from Kuroki-Arita\cite{kuroki}, a {\it gapful} $\phi_s$  
does exist, although its $\lambda$ is slightly 
smaller than that for the d-wave.   
We have called the former solutions `d-wave' in that the 
gap function changes sign as $+-+-$ azimuthally 
(i.e., within each band), while the 
latter gap `s-wave' in that it does not.  

As a hallmark that the scattering of the (intra-band) pairs 
across the inter-band nesting is exploited, all these solutions indeed have 
\begin{equation}
\phi_{{\rm AA}}(k)\phi_{{\rm BB}}(k') < 0.  
\end{equation}
We can confirm in Eliashberg's eq.(\ref{eliashberg}) that 
the sign change across the two bands works favorably 
in increasing $\lambda$ if we note 
$V_{\alpha\beta}^{(2)}(k-k')>0$ 
(peaked around ${\bf k}-{\bf k'}=(0,0,\pi)$) and a relation 
for multiband Green's function 
$G_{\alpha\beta}(-k)=G_{\alpha\beta}^*(k)$ 
which gives $G_{AB}(k')G_{AB}^*(-k')>0$. 

\begin{figure}
\begin{center}
\leavevmode\epsfysize=40mm 
\epsfbox{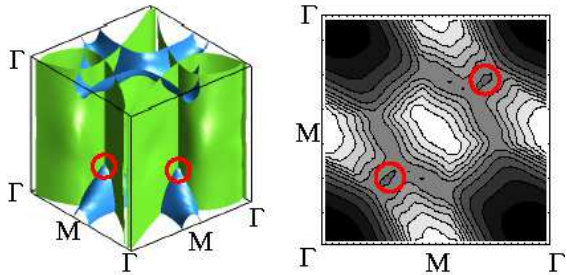}
\caption{Left panel superposes $\phi_{d_1}$ and $\phi_{d_2}$ 
for the bonding band. Right panel plots 
$\left|\phi_{d_1}+i\phi_{d_2}\right|$ for $k_z=0$. 
The red circles denote point nodes in $\phi_{d_1}+i\phi_{d_2}$.}
\label{d+id}
\end{center}
\end{figure}

The fact that the dominant `d-wave' solutions 
are doubly degenerate can be understood by a group theoretical 
argument\cite{sigrist-ueda}, in which 
these solutions belong to $\Gamma_6^+$ representation 
for the honeycomb system (while `s-wave' to $\Gamma_1^+$). 
The true gap function below $T_c$ to maximize the gap should be a 
linear combination of the 
two d-waves, 
\begin{equation}
\phi_{d_1}+i\phi_{d_2}, 
\end{equation}
which breaks the time-reversal symmetry.  
This combination has {\it point nodes} on the Fermi 
surface, which is curious but natural 
as evident from Fig.\ref{d+id} which superposes 
two sets of nodal planes to show how 
the nodal planes intersect each other along some lines for 
the doubly degenerate function and how these lines in turn intersect 
the closed Fermi surface.  

The reason why `s-wave' is only subdominant may be traced back 
to the Fermi surface, which is rather extended in 
${\bf k}$ space in this particular case, 
so that there is an appreciable contribution 
from the intra-band pair scattering 
that the d-wave can exploit.

\begin{figure}
\begin{center}
\leavevmode\epsfysize=55mm 
\epsfbox{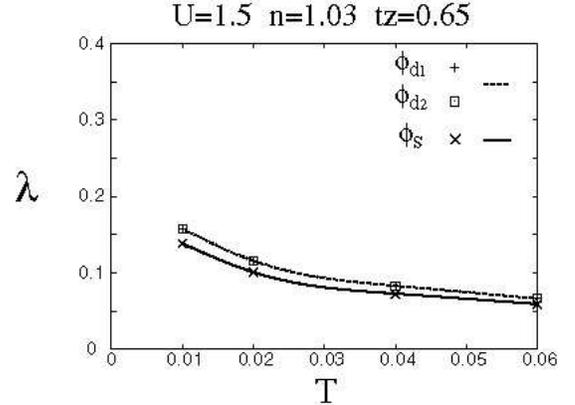}
\caption{The eigenvalue of Eliashberg's equation $\lambda$ 
versus temperature. The largest solutions are doubly degenerate. 
Lines in this and following figures are guide to the eye. }
\label{su-MgB2}
\end{center}
\end{figure}

If we turn to the temperature dependence 
of $\lambda$ in Fig.\ref{su-MgB2}, we can see that $\lambda$ is 
significantly smaller than unity 
even for $T\rightarrow 0.01|t|$.  This implies 
that the spin fluctuation alone is not strong enough to 
realize the d-wave superconductivity in this temperature range 
for the band filling and $t_z$ taken here. 

On the other hand, the $\lambda$ for the gapful `s-wave' pairing 
is seen to have nearly the same magnitude as that of the gapless d-wave, 
although $\lambda$ is again small.  
Thanks to the absence of nodes on the Fermi surface, 
this one has a gapful pairing, 
which is eligible for assisting the phonon-mediated pairing 
if the electron-phonon interaction is considered on top of the 
electron-electron interaction\cite{comment}.
So we conclude that inter-band spin fluctuations can work cooperatively 
with intra-band phonons to realize a gapful superconductivity.

\subsection{Optimization for the 3D honeycomb lattice}
Let us depart from MgB$_2$ to 
move on to the strong coupling regime in search of 
superconductivity from electron repulsion alone.  
In general, pairing instability mediated by
spin fluctuations in 3D systems is definitely weaker
than that in 2D systems\cite{arita,Monthoux}. 
This has been shown in a FLEX study for the 
Hubbard model by three of the present authors\cite{arita}, who have 
identified its reason in the $k$ space volume fraction of the
effectively attractive pair scattering region that is 
much smaller in 3D than in 2D.
Furthermore, 3D systems have a strong tendency 
toward various magnetic orders,
so that to identify the 3D systems that favor 
superconductivity from electron repulsion 
becomes a challenging problem.

\begin{figure}
\begin{center}
\leavevmode\epsfysize=40mm 
\epsfbox{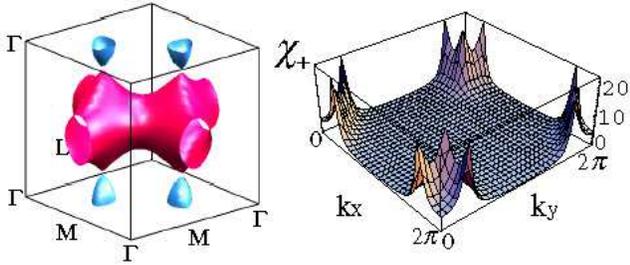}
\caption{Fermi surface (left) and $\chi_+(k_z=\pi)$ (right) 
for the optimized parameter set of $U=8, n=1.15, t_z=0.7$ 
at $T=0.01$, where the pairing symmetry is d-wave.}
\label{U8-n1.15-tz0.7-fermi-chi}
\end{center}
\end{figure}

\begin{figure}
\begin{center}
\leavevmode\epsfysize=55mm 
\epsfbox{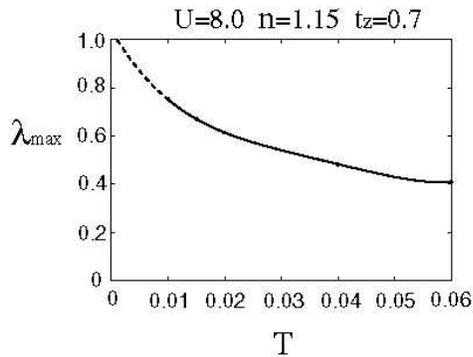}
\caption{$\lambda$ versus temperature for the optimized parameter set 
employed in the previous figure. Dotted line 
is a spline extrapolation to lower temperatures.}
\label{su-U8-n1.15-tz0.7}
\end{center}
\end{figure}

Here we have optimized the pairing instability
in the layered honeycomb lattice by varying the inter-layer 
hopping $t_z$ and the band filling $n$.  
Namely, we have searched for sets of parameter values that give
large values of $\lambda$ {\it without} encountering antiferromagnetic
instability at low temperatures.
The resulting best parameter set is found to be 
$U=8, n=1.15, t_z=0.7$ (inset of Fig.\ref{U8-n1.15-tz0.7-fermi-chi}), 
for which
the Fermi surface and the (inverse) spin susceptibility are shown in
Fig.\ref{U8-n1.15-tz0.7-fermi-chi}. In Fig.\ref{su-U8-n1.15-tz0.7} we
can see that the 
estimated $T_c\sim 0.001$ ($\lambda_{{\rm max}}\rightarrow 1$). 
The pairing symmetry is again $\phi_{d_1}+i\phi_{d_2}$. 

\subsection{Effect of the inter-band nesting}
\begin{figure}
\begin{center}
\leavevmode\epsfysize=40mm 
\epsfbox{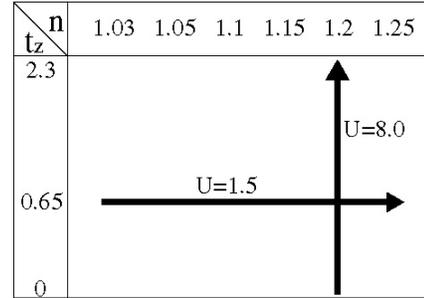}
\caption{The paths we have focused on in the present study 
in the parameter space of the inter-layer 
hopping $t_z$ and the band filling $n$.}
\label{diagram}
\end{center}
\end{figure}

\begin{figure}
\begin{center}
\leavevmode\epsfysize=85mm 
\epsfbox{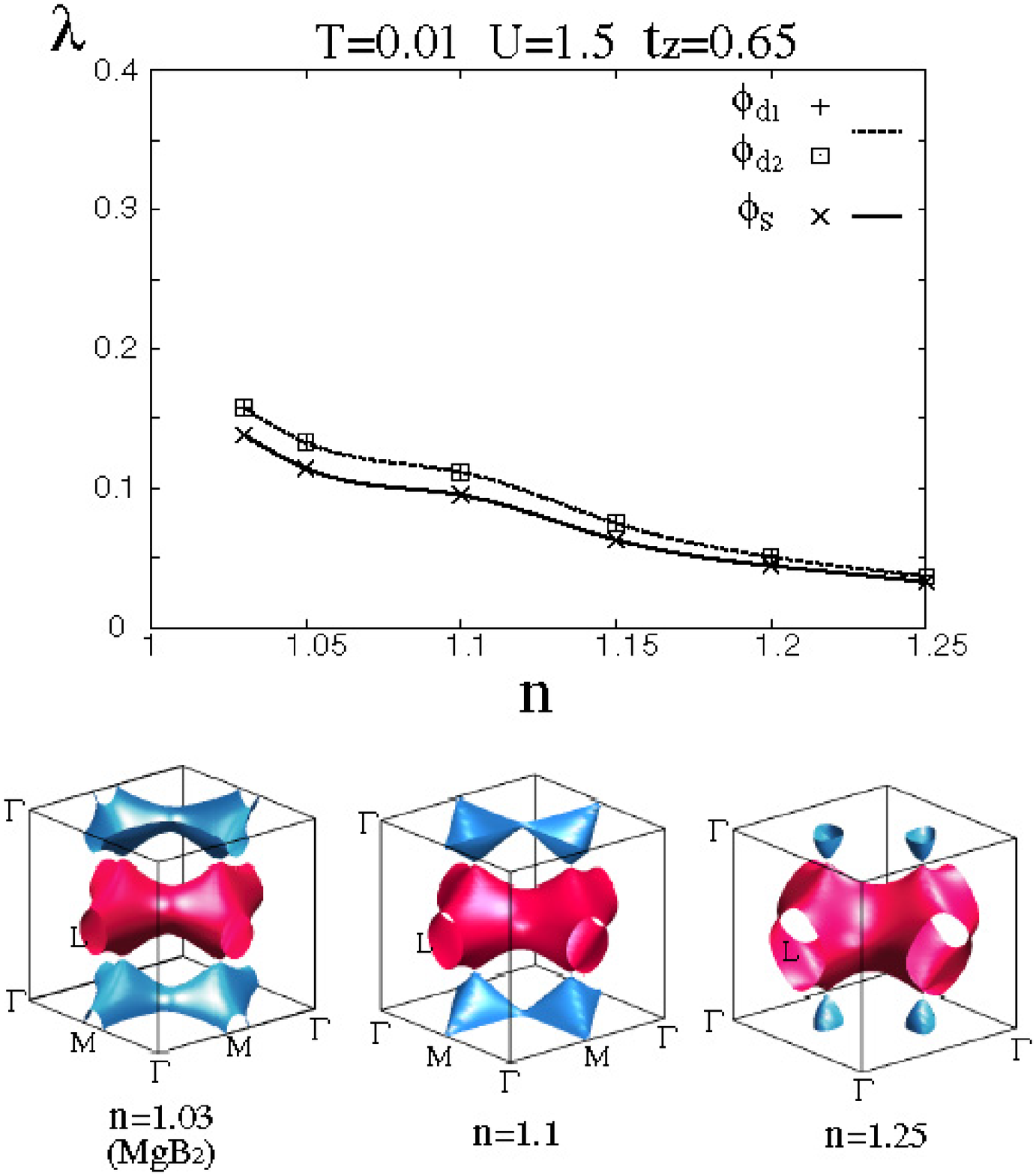}
\caption{$\lambda$ versus $n$ for $U=1.5,t_z=0.65,T=0.01$. The 
d-wave solutions having the largest $\lambda$ are doubly degenerate.}
\label{su-U1.5-tz0.65}
\end{center}
\end{figure}

Apart from the above optimization, we have systematically explored 
how the inter-band nesting affects the pairing 
in the 3D honeycomb lattice.
Since we wish to separate out the effect of the 
band filling and the hopping in the $z$ direction, 
we have done this along two paths displayed in 
Fig.\ref{diagram}. The first path starts from the parameter values 
corresponding to the $p\pi$ bands in MgB$_2$, 
while the second path includes the 
parameter regime ($t_z<0.2$) which corresponds to GIC except 
for the value of $U$.
 The result along the first path
is shown in Fig.\ref{su-U1.5-tz0.65}, where the $n$-dependence 
of $\lambda_{{\rm max}}$ is plotted for $U=1.5,t_z=0.65$ at $T=0.01$.
We can see that the pairing instability becomes weaker as $n$ is 
increased, which is natural since the inter-band
nesting becomes degraded along this path.

\begin{figure}
\begin{center}
\leavevmode\epsfysize=85mm 
\epsfbox{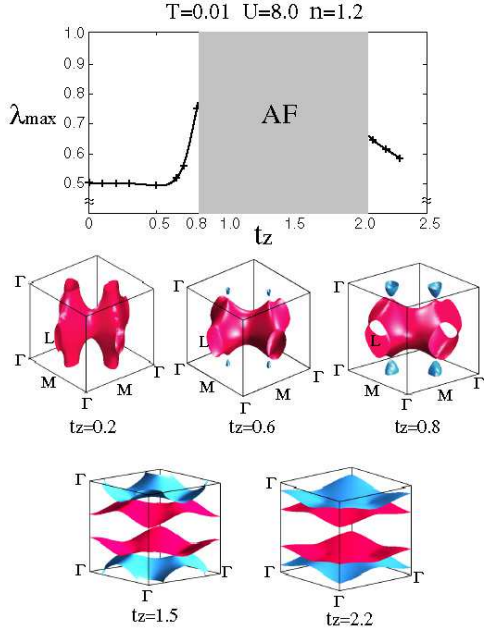}
\caption{$\lambda$ versus $t_z$ for
 $U=8,n=1.2,T=0.01$.  Grey region indicates the 
 antiferromagnetic phase.  
 The insets depict the shape of Fermi surface at 
 five points on the horizontal axis.}
\label{su-U8-n1.2}
\end{center}
\end{figure}

The result along the second path is displayed 
in Fig.\ref{su-U8-n1.2}, where $\lambda$ is plotted as a function of $t_z$ for
$n=1.2$ at $T=0.01$.  
Here we have adopted a rather large 
$U=8$, because we want to have sizeable $\lambda$ 
over a wide range of $t_z$ including 
the case of bad nesting.  
There, we have covered 
both the layered case ($t_z<1$) and a quasi-1D case ($t_z>1$).  
As indicated in Fig.\ref{su-U8-n1.2}, 
antiferromagnetism occurs (i.e., $\chi \rightarrow \infty$) 
when the inter-band nesting becomes too strong as $t_z\rightarrow 1$ 
both from below and from above. 
Before the transition to antiferromagnetism occurs 
$\lambda$ increases both from below and from above, 
indicating that inter-band 3D nesting 
is effective, the layered case $(t_z=0.8)$ is 
more advantageous than the case of 2D $(t_z=0)$ and quasi-1D $(t_z>1)$ on the present path. 

\section{conclusion}
In conclusion, we have studied the possibility of 
spin-fluctuation mediated superconductivity in 3D honeycomb 
lattice systematically. We have shown that if we take the parameter 
set corresponding to the $p\pi$ bands in MgB$_2$, 
the spin fluctuation favors the gapful pairing, which 
suggests that the electron correlation can help the
phonon in forming the Cooper pairing. 
Experimentally, the electron repulsion acting 
constructively may be confirmed if 
some phase-sensitive method can detect the gap function having 
opposite signs in two $\pi$-bands.
When strong enough, the electron repulsion alone 
will give rise to 
a $d$-wave pairing, with a time-reversal broken
 $\phi_{d_1}+i\phi_{d_2}$ symmetry associated with 
 degenerate representations in the non-Bravais lattice 
 with peculiar point nodes on the Fermi surface. 
\section{acknowledgements}
Numerical calculations were performed at the Computer Center 
and the ISSP Supercomputer Center of University
 of Tokyo. This study is in part supported by a Grant-in-aid for scientific
 research from the Ministry of education of Japan.

\end{multicols}

\begin{references}
\bibitem{U=1.5} $U/|t|<2$ is appropriate here because $t\approx 1.7$ eV
[K. Voelker, V. I. Anisimov, and T. M. Rice, cond-mat/0103082]
for $p_\pi$ orbitals in MgB$_2$,
while $U$ for 2p orbitals should be less than $\sim 4$ eV 
as estimated for the O2p orbitals in the high-$T_C$ cuprates.
\bibitem{comment} Here we assume that phonons have long wavelength, 
so that they mainly mediate attractive pairing interaction within 
each band.
\end{references}
\end{document}